\documentclass{aa}  

\usepackage{graphicx}
\usepackage{amsmath}
\usepackage{amssymb}
\usepackage{txfonts}
\usepackage[colorlinks=true,linkcolor=blue,citecolor=blue,urlcolor=blue]{hyperref}

\newcommand{\hi}{H\,{\sc i}}

\newcommand{\cii}{C\,{\sc ii}}

\newcommand{\oi}{O\,{\sc i}}
\newcommand{\omg}{$\Omega_{\rm Z,ISM}$}

\usepackage{color, colortbl} % For color font during drafting

\usepackage{xspace} % For correct spaces after new commands %courtesy Darach
\begin{document} 

\title{Gauging the mass of metals in the gas phase of galaxies from the Local Universe to the Epoch of Reionization}
\titlerunning{The ISM metal mass of galaxies through cosmic time}
%\title{A massive, metal-rich galaxy at $z=8.685$}
%\subtitle{Clues to the assembly and chemical enrichment of the first galaxies}
%\titlerunning{A massive, metal-rich galaxy at $z=8.685$}

\author{
K. E. Heintz\inst{1,2} 
\and A.~E.~Shapley\inst{3}
\and R.~L.~Sanders\inst{3,4}
\and M.~Killi\inst{1,2}
\and D.~Watson\inst{1,2}
\and G.~Magdis\inst{1,5,2}
\and F.~Valentino\inst{1,2}  
%\newauthor
\and M.~Ginolfi\inst{6}
\and D.~Narayanan\inst{7,1} 
\and T.~R.~Greve\inst{1,5}
\and J.~P.~U.~Fynbo\inst{1,2}
\and D.~Vizgan\inst{8,1}
\and S.~N.~Wilson\inst{1,2} 
}
\institute{
Cosmic Dawn Center (DAWN), Denmark 
\and
Niels Bohr Institute, University of Copenhagen, Jagtvej 128, DK-2200 Copenhagen N, Denmark 
\and
Department of Physics and Astronomy, University of California, Los Angeles, 430 Portola Plaza, Los Angeles, CA 90095, USA 
\and
Department of Physics, University of California, Davis, 1 Shields Avenue, Davis, CA 95616, USA 
\and
DTU-Space, Technical University of Denmark, Elektrovej 327, 2800 Kgs. Lyngby, Denmark 
\and
Dipartimento di Fisica e Astronomia, Università di Firenze, Via G. Sansone 1, 50019, Sesto Fiorentino (Firenze), Italy
\and
Department of Astronomy, University of Florida, Gainesville, FL 32611, USA 
\and
Department of Astronomy, University of Illinois at Urbana-Champaign, 1002 West Green St., Urbana, IL 61801, USA
}
\authorrunning{Heintz et al.}

\date{Received \today; accepted \today}

% \abstract{}{}{}{}{} 
% 5 {} token are mandatory
\abstract{ 
The chemical enrichment of dust and metals are vital processes in constraining the star formation history of the universe. These are important ingredients in the  formation and evolution of galaxies overall. Previously, the dust masses of high-redshift star-forming galaxies have been determined through their far-infrared continuum, however,  equivalent, and potentially simpler, approaches to determining the metal masses have yet to be explored at $z\gtrsim 2$.
Here, we present a new method of inferring the metal mass in the interstellar medium (ISM) of galaxies out to $z\approx 8,$ using the far-infrared [\cii]$-158\mu$m emission line as a proxy. We calibrated the [\cii]-to-$M_{\rm Z,ISM}$ conversion factor based on a benchmark observational sample at $z\approx 0$, in addition to gamma-ray burst sightlines at $z>2$ and cosmological hydrodynamical simulations of galaxies at $z\approx 0$ and $z\approx 6$. We found a universal scaling across redshifts of $\log (M_{\rm Z,ISM}/M_\odot) = \log (L_{\rm [CII]}/L_\odot) - 0.45,$ with a 0.4 dex scatter, which is constant over more than two orders of magnitude in metallicity. We applied this scaling to recent surveys for [\cii] in galaxies at $z\gtrsim 2$ and compared their inferred $M_{\rm Z,ISM}$ to their stellar mass ($M_\star$). In particular, we determined the fraction of metals retained in the gas-phase ISM, $M_{\rm Z,ISM} / M_\star$, as a function of redshift and we showed that an increasing fraction of metals reside in the ISM of galaxies at higher redshifts. We  place further constraints on the cosmic metal mass density in the ISM (\omg) at $z\approx 5$ and $\approx 7$ based on recent estimates of the [\cii]$-158\mu$m luminosity functions at these epochs, yielding $\Omega_{\rm Z,ISM} = 6.6^{+13}_{-4.3}\times 10^{-7}\,M_\odot\, {\rm Mpc}^{-3}$ ($z\approx 5$) and $\Omega_{\rm Z,ISM} = 2.0^{+3.5}_{-1.3}\times 10^{-7}\,M_\odot\, {\rm Mpc}^{-3}$ ($z\approx 7$), respectively. These results are consistent with the expected metal yields from the integrated star formation history at the respective redshifts. This suggests that the majority of metals produced at $z\gtrsim 5$ are confined to the ISM, with strong implications that disfavor efficient outflow processes at these redshifts. Instead, these results  suggest that the extended [\cii] halos predominantly trace the extended neutral gas reservoirs of high-$z$ galaxies.
}


\maketitle
%
%-------------------------------------------------------------------

\section{Introduction} \label{sec:intro}

The metal enrichment of the interstellar medium (ISM) of a galaxy is an imprint left behind by stellar processing. It encodes information about the star formation history (SFH) of the galaxy and simultaneously provides insight into the complex processes that regulate the metal content, such as large-scale gas infall and outflows \citep{Heintz23b}. Similarly, the amount of dust directly informs the dust yields from the explosions of core-collapse supernovae \citep[e.g.,][]{Gall14,Lesniewska19} and the efficiency of metals depleting into dust grains through ISM growth \citep[e.g.,][]{Dwek98}. Therefore, obtaining a complete census of the dust and metals in galaxies, particularly at the earliest epochs, is vital to constraining these processes and understanding the formation and evolution of the first generation of galaxies overall.

Early attempts to gauge the metal census at high redshift reported that the bulk of the expected metals from the integrated SFH at $z\approx 2$ were missing from the ISM, having likely been expelled through outflows \citep{Pettini99,Prochaska03,Ferrara05,Bouche07}. This led to the long-standing ``missing metals problem,'' of which only 30\% to 90\% of the expected cosmic density of metals could be accounted for in the intergalactic medium (IGM), the ISM, and in stars \citep{Ferrara05,Bouche07}. However, these early studies were significantly affected by biased selections of quasars and, as a consequence, they were unable to identify the most metal- and dust-rich foreground absorbers \citep[e.g.,][]{Fall93,Heintz18,Krogager19}. Accounting for this bias and the total amount of metals locked into dust grains has yielded cosmic metal mass densities in the ISM of galaxies at $z\gtrsim 2.5,   $ which is consistent with the total integrated SFH metal yield \citep{Peroux20}. Yet even this approach is limited in the sense that quasar absorbers are increasingly less robust tracers of the ISM in galaxies at higher redshifts, mainly probing  the diffuse circumgalactic medium in the outskirts of their absorbing galaxies \citep{Neeleman19,Stern21,Heintz22}. Furthermore, they are virtually impossible to detect beyond $z\approx 5$ due to the near-complete suppression of the emission located in the Lyman-$\alpha$ forest caused by the Gunn-Peterson effect. It is thus imperative that we establish a complementary approach to constrain the metal mass in the ISM of early galaxies. 

While several methods exist to determine the dust mass \citep[e.g.,][]{Dwek98,Draine07,Scoville14,Sommovigo21,Sommovigo22} and the metallicity through nebular emission strong-line ratios \citep[e.g.,][]{Maiolino08,Kewley19,Sanders21} of the ISM of individual galaxies, there is currently no simple way to directly measure the total metal mass. The ISM metal mass, $M_{\rm Z,ISM}$, has previously been measured through a combination of the metallicity and the gas mass \citep{Sanders23a,Eales23}, but this approach has been limited by the difficulty of deriving metallicities from optical nebular emission lines from ground-based facilities beyond $z\approx 3$ \citep[though see recent efforts based on joint JWST and ALMA observations, e.g.,][]{Heintz23a}. In this paper, we present a novel approach to infer the total ISM metal mass of individual galaxies using only the [\cii]$-158\mu$m line luminosity as a proxy. The [\cii]$-158\mu$m emission is advantageous due to its immense brightness as one of the strongest ISM cooling lines \citep{Hollenbach99,Wolfire03,Lagache18}, and it has efficiently been used as a viable tracer of cold gas in both local and distant universe \citep[e.g.,][]{Stacey10,Madden97,Madden20,Cormier15,Zanella18,DessaugesZavadsky20,Heintz21,Heintz22,Vizgan22a,Vizgan22b,Liang23}. 

There are several pieces of evidence that point to [\cii] being a potential effective tracer of the total ISM metal mass in galaxies. Firstly, carbon is the second most abundant metal by mass in the universe (following oxygen). Secondly, the ionization potential of neutral carbon (IP = 11.26 eV) is sufficiently below that of neutral hydrogen, such that the majority of carbon will be in the singly ionized state in the neutral ISM. Furthermore, since [\cii] has been observed to originate from the multiple phases of the ISM, from the outskirts of molecular clouds to the neutral and ionized ISM \citep{Pineda14,Vallini17,Pallottini19,RamosPadilla22}, the emission from [\cii] will probe metals in a large range of physical environments and gas properties. Finally, the observed anti-correlation with metallicity of the [\cii]-to-\hi\ abundance ratio \citep{Heintz21,Vizgan22b,Liang23} points to a constant scaling between $L_{\rm [CII]}$ and $M_{\rm Z,ISM}$, as also noted for other line emission gas tracers \citep{Eales23}. 

To gauge the robustness and the applications of [\cii] as a proxy for $M_{\rm Z,ISM}$, we have structured the paper as follows. First, we provide an overview of the compiled observational samples, the adopted simulations, and the overall methodology to derive the [\cii]-to-$M_{\rm Z,ISM}$ calibration in Sect.~\ref{sec:data}. Then we present our results in Sect.~\ref{sec:res}, focusing on the $M_{\rm Z,ISM}$ to stellar mass content as a function of redshift, and in Sect.~\ref{sec:rhomet}, we attempt to constrain the cosmic ISM metal mass density \omg\ at $z\gtrsim 5$. We summarize our conclusions in Sect.~\ref{sec:conc}.

Throughout the paper, we adopt the concordance $\Lambda$CDM cosmological model with $\Omega_{\rm m} = 0.315$, $\Omega_{\Lambda} = 0.685$, and $H_0 = 67.4$\,km\,s$^{-1}$\,Mpc$^{-1}$ \citep{Planck18}. We assume the initial mass function (IMF) from \citet{Chabrier03}
 and solar abundances from \citet{Asplund09}, with $Z_\odot = 0.0134$.

%%%%%%%%%%%%%%%%%%%%%%%%%%%%%%%%%%%%%%%%%%%%%%%%%%%%%%%%%%%

\section{Observational samples, simulations, and methods} \label{sec:data}

In the following sections, we compile all the observational data and simulations used to measure the benchmark ISM metal mass, $M_{\rm Z,ISM}$, and to calibrate the [\cii]-to-$M_{\rm Z,ISM}$ conversion factor in galaxies.

\subsection{Observational galaxy sample at \texorpdfstring{$z\approx 0$}{z~0}}

\begin{figure}%[t!]
\centering
\includegraphics[width=9cm]{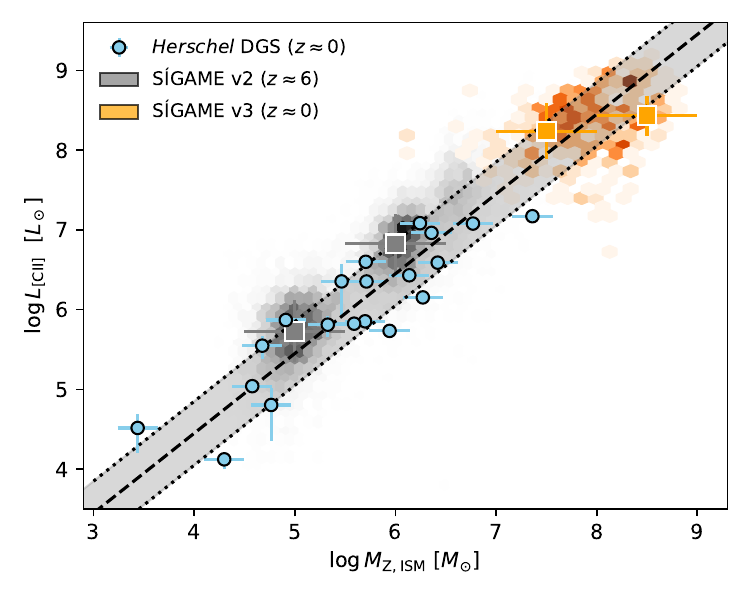}
\caption{Metal mass, $M_{\rm Z,ISM}$, vs. [\cii] luminosity, $L_{\rm [CII]}$. The observed galaxy samples at $z\approx 0$ (see text) with direct measurements of $M_{\rm Z,ISM}$ and $L_{\rm [CII]}$ are shown by the blue circles. The grey- and orange-shaded 2D hexagonal histograms represent simulated galaxies from {\tt SÍGAME}-v2 ($z\approx 6$) and v3 ($z\approx 0$), and their mean and $1\sigma$ distributions are marked by the grey and orange squares respectively. The dashed line indicates the best-fit relation $\log (M_{\rm Z,ISM}/M_\odot) = \log (L_{\rm [CII]}/L_\odot ) - 0.45$ between all data sets and the grey-shaded region indicates the 0.4\,dex RMS scatter.}
\label{fig:lciimmet}
\end{figure}

We first considered the observational sample of galaxies at $z\approx 0$ from the {\it Herschel} Dwarf Galaxy Survey \citep[{\it Herschel} DGS;][]{Madden13}, for which measurements of the metallicities \citep[][mainly using the $R_{23}$ method from \citealt{Pilyugin05}]{Madden13}, \hi\ gas masses \citep{RemyRuyer14}, and [\cii] luminosities \citep{Cormier15} for a subset of the ``compact sample'' have been derived. This sample includes galaxies with metallicities in the range $12+\log{\rm (O/H)} = 7.2 - 8.4$ (i.e. $\log Z/Z_\odot = -1.5$ to $-0.3$, assuming $12+{\rm (O/H)_\odot} = 8.69$; \citealt{Asplund09}) and \hi\ gas masses $M_{\rm HI} = 2\times 10^{6} - 3.5\times 10^{9}\,M_\odot$. These low-metallicity, gas-rich galaxies resemble the typical high-$z$ galaxy population and therefore serves as an ideal benchmark. For these galaxies we derive the metal mass, $M_{\rm Z,ISM}$, as
\begin{equation}
    M_{\rm Z,ISM} = M_{\rm HI} \times 10^{12+\log{\rm (O/H)-8.69}} \times Z_\odot, 
    \label{eq:mmets}
\end{equation}
assuming solar abundance patterns, where $12+\log{\rm (O/H)}_\odot = 8.69$ is the solar oxygen abundance and $Z_\odot = 0.0134$ is the solar metallicity by mass. We only consider the \hi\ gas since it reflects the dominant ISM gas mass contribution in both local \citep{Leroy09,Morselli21} and high-$z$ \citep{Scoville17,Heintz21,Heintz22} galaxies, and to be consistent with the GRB measurements (see Sect.~\ref{ssec:ciimz}). Further, the majority of metals by mass are expected to be associated with the neutral gas-phase with only minor contributions from molecular regions. We derive metal masses in the range $M_{\rm Z,ISM} = 2.7\times 10^{3} - 2.3\times 10^{7}\,M_\odot$ for this benchmark sample of galaxies at $z\approx 0$. The results are shown in Fig.~\ref{fig:lciimmet}. We note that recent efforts to compile a more extensive benchmark local galaxy sample have been presented by \citep{Ginolfi19,Hunt20}. However, this sample (so far) lacks comparable [\cii] detections, so we do not further consider it  in this work. 

\subsection{Simulations of galaxies at \texorpdfstring{$z\approx 0,6$}{z~0,6}}

To support the observational data, we further include the recent simulations of galaxies through the Simulator of Galaxy Millimeter/submillimeter Emission ({\tt SÍGAME}) framework \citep{Olsen17}\footnote{https://kpolsen.github.io/SIGAME/index.html}. This simulation provides detailed modelling of the far-infrared line emission from galaxies extracted from the particle-based cosmological hydrodynamics simulation {\tt Simba} \citep{Dave19}. We consider the results derived for galaxies at $z\approx 6$ as part of the {\tt SÍGAME} version 2 (v2) presented by \citet{Leung20} and \citet{Vizgan22a}, and the more recent v3 results applicable to galaxies at $z\approx 0$ \citep{Olsen21}. To derive the metal masses for these simulated sets of galaxies, we used Eq.~\ref{eq:mmets} above. Following \citet{Vizgan22b}, we represent $M_{\rm HI}$ by the total ``diffuse'' \hi\ component in the {\tt SÍGAME}-v2 simulations, which is equal to the sum of its ionized and atomic hydrogen gas mass and distinct from the ``dense'' fraction of the mass of each fluid element, and extract directly the \hi\ component from the {\tt SÍGAME}-v3 model. For both simulations, we adopted the star formation rate weighted gas-phase metallicity, $Z_{\rm SFR}$, which best represent the metallicities of the H\,{\sc ii} regions inferred through the emission-line measurements.

%%%%%%%%%%%%%%%%%%%%%%%%%%%%%%%%%%%%%%%%%%%%%%%%%%%%%%%%%%%
%%%%%%%%%%%%%%%%%%%%%%%%%%%%%%%%%%%%%%%%%%%%%%%%%%%%%%%%%%%

\subsection{The \texorpdfstring{[\cii]-to-$M_{\rm Z,ISM}$}{[CII]-M(Z,ISM)} calibration} \label{ssec:ciimz}

We compared the results from the simulated data sets to the observational dwarf galaxy sample at $z\approx 0$ in Fig.~\ref{fig:lciimmet}. The {\tt SÍGAME}-v2 simulations at $z\approx 6$ best represent the observed data, as they cover a similarly broad range in the [\cii] luminosities of $L_{\rm [CII]} = 6.6\times 10^{3} - 8.1\times 10^{8}\,L_{\odot}$ and metal masses, $M_{\rm Z,ISM} = 1.8\times 10^{4} - 1.0\times 10^{8}\,M_\odot$. The {\tt SÍGAME}-v3 simulations mainly represent the high-$L_{\rm [CII]}$, high-$M_{\rm Z,ISM}$ parameter space, covering $L_{\rm [CII]} = 5.3\times 10^{6} - 1.6\times 10^{9}\,L_{\odot}$, and $M_{\rm Z,ISM} = 1.1\times 10^{6} - 6.7\times 10^{8}\,M_\odot$. We determined the correlation between $L_{\rm [CII]}$ and $M_{\rm Z,ISM}$ by first computing the Pearson $r$ and $p$ correlation coefficients using {\tt scipy}'s {\tt stat} module. For the simulations datasets, we considered the mean and standard deviations in bins of the data, as visualized in Fig.~\ref{fig:lciimmet}. This yields $r = 0.93$ and $p=7.6\times 10^{-11}$, suggesting highly correlated data. Then, we used the {\tt linear regression} module in {\tt Scikit} to estimate the linear best-fit relation, in addition to the root mean square (RMS) and the $r^2$ scatter of the data. We found a slope consistent with unity (formally $0.91\pm 0.10$), over more than five orders of magnitudes in $L_{\rm [CII]}$ and $M_{\rm Z,ISM}$,   with a unique constant ratio of
\begin{equation}
    \log (M_{\rm Z,ISM}/M_\odot) = \log (L_{\rm [CII]}/L_\odot ) - 0.45\pm 0.40,
    \label{eq:ciimmet}
\end{equation}
where the uncertainty represents the RMS scatter, with an $r^2$ value of 0.85.

%due to the apparent universal $L_{\rm [CII]}-$SFR relation \citep[with a scatter of $\approx 0.4$\,dex;][]{DeLooze14}, consistent from local to high-redshift galaxies \citep{DeLooze14,Schaerer20,Romano22}, and the connection between $L_{\rm [CII]}-M_{\rm HI}$ with metallicity \citep{Heintz21}, these quantities represent an equivalent $L_{\rm [CII]}-M_{\rm HI}-Z$ or SFR$-M_{\rm HI}-Z$ fundamental plane relation similar to the one observed in local galaxies \citep{Bothwell13,Curti20}. 
%We observe, however, no equivalent unique trend in the compiled observational and simulated data set for the connection between $L_{\rm [OIII]}$ and $M_{\rm Z,ISM}$, where [\oiii] is also thought to trace star formation. This is likely because the [\oiii]$-88\mu$m emission purely originates within the central, most intense \hii\ regions, and is thus more sensitive to the degree of ionization than the amount of metals, with an ionization potential of IP$_{{\rm O}^{+}} = 35.1$\,eV compared to IP$_{\rm C} = 11.2$\,eV. 

%%%%%%%%%%%%%%%%%%%%%%%%%%%%%%%%%%%%%%%%%%%%%%%%%%%%%%%%%%%
\begin{figure}%[!t]
\centering
\includegraphics[width=9cm]{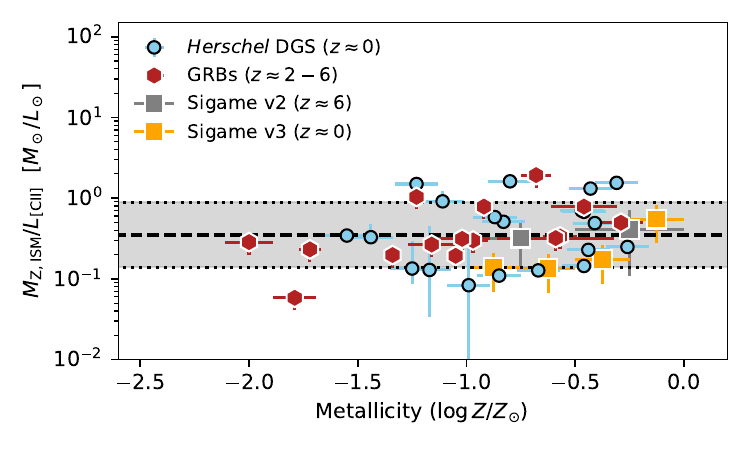}
\caption{$M_{\rm Z}-L_{\rm [CII]}$ relation as a function of metallicity. Red hexagons show the inferred line-of-sight measurements from GRB sightlines (see text for details), the blue circles denote the observed galaxy samples at $z\approx 0$ with direct measurements of $M_{\rm Z,ISM}$, and the average values of $M_{\rm Z}/L_{\rm [CII]}$ in bins of 0.5 dex in metallicity predicted by the simulations at $z\approx 0$ are shown as orange squares and at $z\approx 6$ as the grey squares. The constant $M_{\rm Z}-L_{\rm [CII]}$ ratio and the RMS scatter of the data are shown by the dashed line and grey-shaded region, respectively.}
\label{fig:mmetlciimet}
\end{figure}
%%%%%%%%%%%%%%%%%%%%%%%%%%%%%%%%%%%%%%%%%%%%%%%%%%%%%%%%%%%

To further test any potential offsets in the $L_{\rm [CII]}-M_{\rm Z,ISM}$ relation, we show this ratio as a function of metallicity in Fig.~\ref{fig:mmetlciimet}. Here, we also include the relative abundance measurements from GRB absorption line spectroscopy derived by \citet{Heintz21}. This approach infers the ``column'' [\cii] luminosity measured in the line of sight from the spontaneous decay of the excited \cii* transition as $L^{c}_{\rm [CII]} = h\nu_{ul}A_{ul} N_{\rm CII^*}$. Here, $N_{\rm CII^*}$ is the column density of the $^2P_{3/2}$ state of C$^+$, and $\nu_{ul}$ and $A_{ul}$ are the frequency and Einstein coefficient, respectively, of [\cii]. We relate this column [\cii] luminosity to the inferred line of sight metal mass, $M^{c}_{Z} = M_{\rm HI} \times 10^{\rm [M/H]_{tot}} \times Z_\odot$, with $M_{\rm HI}$ being the \hi\ column mass inferred from the column density as $M_{\rm HI} = m_{\rm HI} \times N_{\rm HI}$ and ${\rm [M/H]_{tot}}$ the total absorption metallicity (equivalent to $\log Z/Z_\odot$). These GRB measurements provide accurate estimates of the $L_{\rm [CII]}-M_{\rm Z,ISM}$ ratios only, in pencil-beam sightlines through their host galaxies \citep[see also][]{HeintzWatson20}. 

In Fig.~\ref{fig:mmetlciimet}, we compare the GRB measurements to the observational and simulated galaxy samples described above. We observe a remarkable agreement, with a mean GRB-inferred ratio of $\log (M_{\rm Z,ISM}/L_{\rm [CII]}) = -0.44\pm 0.35$ (error denoting $1\sigma$). Moreover, the GRB sightlines probe galaxies in a large redshift range, $z\sim 2 - 6$, and expand the metallicity regime for which we can determine the $L_{\rm [CII]}-M_{\rm Z,ISM}$ relation, reproducing the constant ratio down to metallicities of $Z/Z_{\odot} = 1\%$. We note, however, that the {\tt SÍGAME}-v3 simulations potentially indicate an increasing $M_{\rm Z,ISM}/L_{\rm [CII]}$ ratio around solar metallicities. This estimate is still within the overall scatter of the relation, yet it may indicate that [\cii] emission is suppressed for a given metal mass in the highest metallicity galaxies. This could be due to more inefficient cooling through the [\cii]$-158\mu$m transition or potentially from lower ionization states. The relation between $L_{\rm [CII]}$ and $M_{\rm Z,ISM}$ derived here is purely an empirical result, based on direct observations and independent simulations, revealing an approximate constant ratio between the two. Crucially, this ratio appears to be constant across redshifts, making the conversion factor universally applicable. 

%show any additional dependencies on e.g. the metallicity

%GRB results.

\section{Results} \label{sec:res}

To apply the [\cii]-to-$M_{\rm Z,ISM}$ conversion factor, we compiled the recent high-$z$ observational samples surveyed for [\cii]: At $z\sim 2,$ we included the observations of main-sequence, star-forming galaxies from \citet{Zanella18}, whereas at $z\sim 4-6,$ we made use of the ALMA Large Program to Investigate C+ at Early Times (ALPINE) survey \citep{LeFevre20,Bethermin20,Faisst20} in addition to the sample of galaxies presented by \citet{Capak15}. At $z\sim 6-8,$ we consider the galaxies from the Reionization Era Bright Emission Line Survey \citep[REBELS;][]{Bouwens22}. In our analysis, we further include the individual measurements of A1689-zD1 \citep[at $z=7.13$;][]{Watson15,Bakx21,Killi23} and S04590 \citep[at $z=8.496$;][]{Heintz23a} since both these high-$z$ galaxies have robust estimates of their ISM gas masses and metallicities. The galaxy samples are all selected to have sufficient auxiliary data to enable derivations of the star formation rate (SFR) and stellar mass ($M_\star$) of each source, and to follow the star-forming galaxy main-sequence at their respective redshifts.

%%%%%%%%%%%%%%%%%%%%%%%%%%%%%%%%%%%%%%%%%%%%%%%%%%%%%%%%%%%
\begin{figure}%[!t]
\centering
\includegraphics[width=9cm]{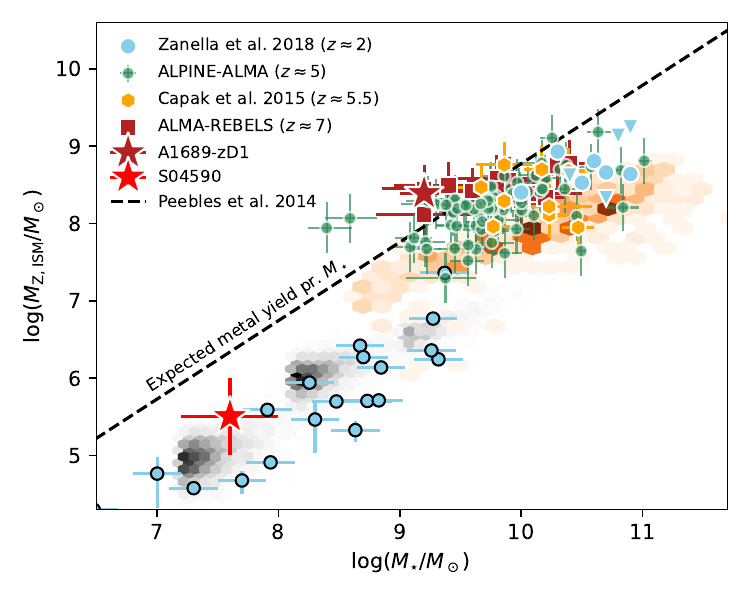}
\caption{Metal mass $M_{\rm Z,ISM}$ as a function of stellar mass $M_\star$ for the compiled high-redshift galaxy sample survey for [\cii]. Colors and symbols denote the respective surveys (see main text for details). The dashed line mark the expected total Type II supernova metal production as a function of $M_\star$ \citep{Peeples14}.}
\label{fig:mstarmet}
\end{figure}
%%%%%%%%%%%%%%%%%%%%%%%%%%%%%%%%%%%%%%%%%%%%%%%%%%%%%%%%%%%

In Fig.~\ref{fig:mstarmet}, we show the inferred metal masses for the compiled sample of high-redshift ($z\gtrsim 2$) galaxies as a function of stellar mass. For all galaxies at $z\gtrsim 2$, except for A1689-zD1 and S04590, we infer the metal mass following the metallicity-independent conversion derived in Eq.~\ref{eq:ciimmet}, $\log (M_{\rm Z,ISM}/M_\odot) = \log (L_{\rm [CII]}/L_\odot) -0.45\pm 0.40$. For A1689-zD1, we infer $M_{\rm Z,ISM}$ based on the derived approximate solar metallicity \citep{Killi23} and the \hi\ gas mass $M_{\rm HI} = 1.8\times 10^{10}\,M_{\odot}$, using the [\cii]-to-\hi\ conversion factor derived by \citet{Heintz21}, which yields $M_{\rm Z,ISM} = 2.45\times 10^{8}\,M_{\odot}$ following Eq.~\ref{eq:mmets}. For S04590, we adopt the metal mass inferred by \citet{Heintz23a} of $M_{\rm Z,ISM}=(3.2\pm 1.5)\times 10^{5}\,M_\odot$, following a similar approach. These estimates are also in agreement with that inferred from the $L_{\rm [CII]}-M_{\rm Z,ISM}$ calibrations for those particular sources. Overall, we find metal masses in the range $M_{\rm Z,ISM} = 2\times 10^{7} - 10^{9}\,M_\odot$, and observe that $M_{\rm Z,ISM}$ generally increases with $M_\star$, which is expected given the increased metal yield for more abundant stellar populations. For comparison, we overplot the $M_{\rm Z,ISM}(M_\star)$ function derived by \citet{Peeples14} from the expected total Type II supernova metal production: 
\begin{equation}\label{eq:mz}
    \log (M_{\rm Z,ISM}/M_\odot) = 1.0146\times \log(M_\star/M_\odot) + \log y + 0.1091,
\end{equation}
based on the star-formation histories by \citet{Leitner12}. Here, $y$ is the nucleosynthetic yields, which we assume to be $y=0.033$ \citep{Peeples14}. We note that these SFHs might be more extended than what is possible for early galaxies at $z\approx 4-7$, simply due to the young age of the universe at these early times. Using more brief SFHs such that the enrichment is dominated by core-collapse supernovae (SNe), \citet{Sanders23a} found that the total metal production is simply proportional to the product of the SNe metal yield, stellar mass, and (1-$R$), where $R$ is the return fraction. This method yields a consistent curve to that derived by \cite{Peeples14}, however, and therefore seems to well represent the expected metal yield also at high-$z$. Generally,  high-redshift galaxies are observed to have $M_{\rm Z,ISM}$ values that are lower than this predicted curve at any given stellar mass. We also note the potentially more significant offset at low stellar masses in Fig.~\ref{fig:mstarmet}. 

%%%%%%%%%%%%%%%%%%%%%%%%%%%%%%%%%%%%%%%%%%%%%%%%%%%%%%%%%%%
\begin{figure}%[!t]
\centering
\includegraphics[width=9cm]{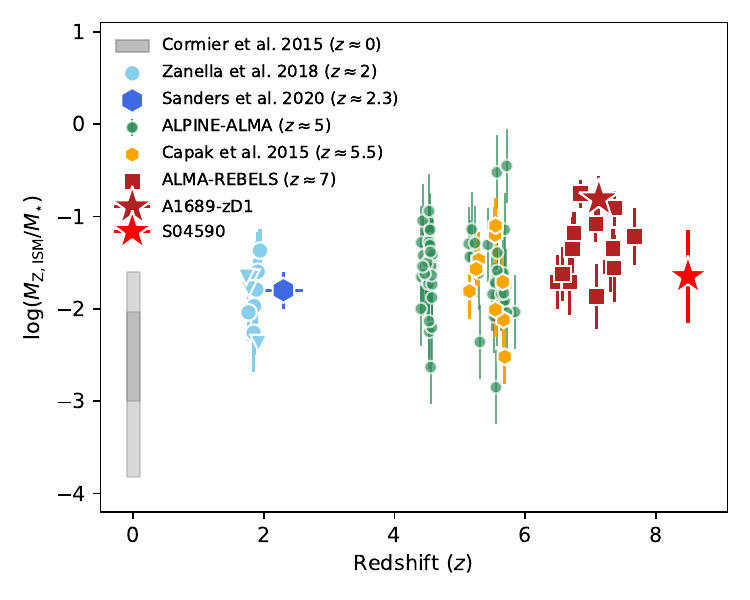}
\caption{Retained metal yield of the ISM, $M_{\rm Z,ISM} / M_\star$, as a function of redshift. Symbol notation follows that of Fig.~\ref{fig:mstarmet}, but now includes the benchmark $z\approx 0$ galaxy sample, marked by dark- and light-gray boxes representing the $1\sigma$ and $2\sigma$ distributions of $M_{Z,ISM} / M_\star$, in addition to the lensed galaxy S04590 from \citet{Heintz23a}. }
\label{fig:mstarmmetz}
\end{figure}
%%%%%%%%%%%%%%%%%%%%%%%%%%%%%%%%%%%%%%%%%%%%%%%%%%%%%%%%%%%

We go on to consider the metal-to-stellar mass ratio, $M_{\rm Z,ISM} / M_\star$, for the compiled sample of galaxies as a function of redshift in Fig.~\ref{fig:mstarmmetz}. This ratio represents how effective galaxies are at retaining metals, given that $M_\star$ is approximately proportional to the total amounts of metals produced as prescribed in Eq.~\ref{eq:mz} \citep[see also][]{Peeples14,Sanders23a}. For the galaxies at $z\sim 0$, $M_{\rm Z,ISM}$ are measured directly through the metallicities and gas masses of the galaxies in the sample. We determine $M_{\rm Z,ISM} / M_\star = (2.5^{+6.6}_{-1.5})\times 10^{-3}$ (median and 16th to 84th percentiles) at $z\approx 0$. This increases to $M_{\rm Z,ISM} / M_\star \approx 10^{-2}$, derived as the average for the sample of galaxies at $z\sim 2$, and further to $M_{\rm Z,ISM} / M_\star \approx 2\times 10^{-2}$ at $z\sim 4-6$ and $M_{\rm Z,ISM} / M_\star \approx 5\times 10^{-2}$ at $z\gtrsim 6$. 
%following approximately $M_{\rm Z,ISM}(z) / M_\star \propto 0.18z$. 
As a reference point, we also include the stacked average of $M_{\rm Z,ISM} / M_\star \approx 10^{-2}$ inferred by \citet{Sanders23a} for galaxies at $z\sim 2-3$. This result is based on rest-frame optical emission lines to infer metallicities and with gas masses inferred from CO, which may introduce a systematic difference compared to our method. Their results indicate that an increasing amount of metals reside in the gas-phase ISM in galaxies at higher redshifts, which also seem to hold for even the most massive systems \citep{Sanders23a}. This likely reflects that feedback mechanisms or outflows are not yet efficient in expelling the metals out of the galaxy ISM at these epochs. On the contrary, most of the metals in local galaxies reside in stars \citep{Peeples14,Muratov17}.

\section{The cosmic metal mass density in galaxies} \label{sec:rhomet}

To further quantify the chemical enrichment of galaxies through cosmic time, we now consider the cosmological metal mass density (\omg) and its evolution with redshift. 
Previously, it has only been possible to infer \omg\ at $z\gtrsim 1$ in quasar absorption-line systems \citep[DLAs, e.g.,][]{Peroux20}, since the metal abundances of high-$z$ galaxies are generally difficult and time consuming to constrain through other approaches such as from nebular line emission and strong-line diagnostics \citep[see e.g.,][for recent reviews]{Kewley19,Maiolino19}. However, there is increasing evidence for DLAs to mainly probe the outskirts of their extended neutral, gaseous halos, in particular at $z\gtrsim 3$ \citep{Neeleman19,Stern21,Yates21,Heintz22}. Therefore, these pencil-beam sightlines do not probe the central star-forming ISM of their galaxy counterparts, so it is imperative to establish alternative approaches to infer \omg\ in galaxies at the highest redshifts. 

\begin{figure}%[!t]
\centering
\includegraphics[width=9cm]{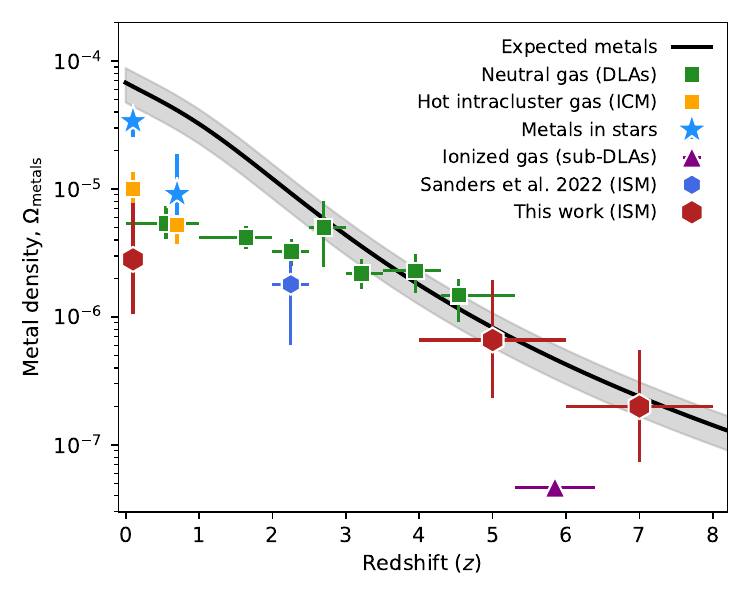}
\caption{Cosmological density of metals as a function of redshift. Red hexagons show the measurements from this work based on the [\cii] luminosity densities at $z\approx 5$ and $\approx 7$, converted to \omg. The other symbols represent the metal densities inferred by various approaches, color-coded as a function of the distinct gas phases they are probing from the compilation of \citet{Peroux20}, including results from \citet{Sanders23a} and with the purple triangle denoting the lower bound derived from \oi\ absorption at $z\approx 6$ in sub-DLAs by \citet{Becker11}. The solid black line shows the expected yield of metals from star formation, assuming $\Omega_{Z}(z) = y\Omega_{\star}(z)$, with $y=0.033$ being the integrated yield of the stellar population and $\Omega_{\star}(z)$ is the stellar mass density quantified by \citet{Walter20}. These measurements suggest that metals mostly reside in the ISM of galaxies at $z\gtrsim 3$.}
\label{fig:rhomet}
\end{figure}

We determined \omg\ based on the derived [\cii]-$158\mu$m luminosity density, defined as $\mathcal{L}_{\rm [CII]} = \int^{\infty}_{L_{\rm [CII]}} L_{\rm [CII]} \phi(L_{\rm [CII]}) dL_{\rm [CII]}$, and the constant [\cii]-to-$M_{\rm Z,ISM}$ scaling derived here. This approach is similar to previous efforts determining the cosmic H$_2$ gas mass density based on the CO luminosity density \citep[e.g.,][]{Walter14,Decarli19,Riechers19} and the \hi\ gas mass density based on an empirical metallicity-dependent scaling to $\mathcal{L}_{\rm [CII]}$ \citep{Heintz21,Heintz22}. Since [\cii] traces the overall star-forming ISM, this approach provides a more direct measure of the mass of metals in the ISM of galaxies, compared to DLA sightlines. We here adopt the [\cii]-$158\mu$m luminosity densities derived by P. Oesch et al. in prep. \citep[see also][]{Heintz22} at $z\approx 5$ and $z\approx 7$, based on the ALMA-ALPINE \citep{Bethermin20} and REBELS \citep{Bouwens22} surveys for [\cii] emission from main-sequence galaxies at the respective redshifts. Integrating the [\cii] luminosity function down to $L_{\rm [CII]} = 10^{7.5}\,L_\odot$, yields luminosity densities of $\log (\mathcal{L}_{\rm [CII]} / L_\odot\, {\rm Mpc}^{-3})= 5.37^{+0.19}_{-0.16}$ (at $z\approx 5$) and $\log (\mathcal{L}_{\rm [CII]} / L_\odot\, {\rm Mpc}^{-3}) = 4.85^{+0.25}_{-0.20}$ (at $z\approx 7$). Applying the [\cii]-to-$M_{\rm Z,ISM}$ calibration, we derive metal mass densities of \omg\ $ = 6.6^{+13}_{-4.3}\times 10^{-7}\,M_\odot\, {\rm Mpc}^{-3}$ and \omg\ $ = 2.0^{+3.5}_{-1.3}\times 10^{-7}\,M_\odot\, {\rm Mpc}^{-3}$, at $z\approx 5$ and $z\approx 7,$ respectively. In Fig.~ \ref{fig:rhomet}, we also show the inferred \omg\ $ = 2.8^{+4.9}_{-1.8}\times 10^{-6}\,M_\odot\, {\rm Mpc}^{-3}$ at $z\approx 0$ based on the [\cii] luminosity function derived by \citet{Hemmati17} at a similar redshift. 

%%%%%%%%%%%%%%%%%%%%%%%%%%%%%%%%%%%%%%%%%%%%%%%%%%%%%%%%%%%
%%%%%%%%%%%%%%%%%%%%%%%%%%%%%%%%%%%%%%%%%%%%%%%%%%%%%%%%%%%

We highlight our measurements in Fig.~\ref{fig:rhomet} as the red hexagons. For comparison, we show the total expected metal yield from stars, defined as $\Omega_{Z,\star}(z) = y\rho_\star(z)/\rho_c$, where $y$ is the integrated yield of the stellar population, $\rho_\star(z)$ the stellar mass density, and $\rho_c$ is the critical density of the universe. Here, we adopt $y=0.033$ from \citet{Peeples14} \citep[see also][]{Peroux20}, the evolutionary function of $\rho_\star(z)$ parametrized by \citet{Walter20}\footnote{Converted to a Chabrier IMF.}, and $\rho_c = 1.26\times 10^{11}\,M_\odot$\,Mpc$^{-3}$ from the concordance $\Lambda$CDM cosmological framework \citep{Planck18}. We find that our measurements are in remarkable agreement with the metal yield predicted by $\Omega_{Z,\star}(z)$ at $z\gtrsim 4$. This suggests that the majority of metals are confined to the galaxy central ISM (as also discussed in Sect.~\ref{sec:res}) and have not yet been expelled to the outer regions through outflows and feedback effects. This suggests that the origin of the extended [\cii] halos are likely not caused by outflows \citep[as previously proposed, e.g.,][]{Maiolino12,Cicone15,Ginfoli20,HerreraCamus21,Akins22}, but supports the scenario where the [\cii] emission instead traces the extended neutral gas reservoirs \citep{Novak19,Novak20,Harikane20,Heintz21,Heintz22,Meyer22}. In this case, there would be some small, but not negligible, in situ star formation in the extended neutral gas disks. Further, these results provide additional evidence for the robustness of the [\cii]-to-$M_{\rm Z,ISM}$ scaling relation derived in this work. At $z\sim 0$, we also find that only $\approx 10\%$ of the expected metal yield resides in the ISM of galaxies, consistent with DLA studies \citep{Peroux20}.
%, and the inferred dominant amount of metals in stars.

In Fig.~\ref{fig:rhomet} we additionally compare our measurements to estimates of \omg\ inferred through various approaches and for different baryonic phases: metals associated with the neutral gas probed via DLAs, metals located in the hot intracluster medium (ICM) and partially ionized gas, as well as the mass of metals attained in stars \citep[see][and references therein]{Peroux20}. We find that our \omg\ estimates inferred from [\cii] are in good agreement with the DLA measurements at $z\gtrsim 3$, showing a consistent decrease in the metal mass density with increasing redshift following that expected from the stellar yield, $\Omega_{Z,\star}(z)$. Direct DLA measurements are, however, only possible out to $z\approx 5$ due to the increasing suppression of the emission located in the Lyman-$\alpha$ forest caused by the Gunn-Peterson effect at these redshifts. \citet{Becker11} attempted to partly alleviate this by measuring the density of \oi\ absorption at $z\approx 6$ in sub-DLAs (probing partly ionized gas down to $N_{\rm HI} = 10^{19}$\,cm$^{-2}$). This point is shown as a lower bound on \omg\ in Fig.~\ref{fig:rhomet} since \omg $\gtrsim \Omega_{\rm OI}$. The method presented here thus provides a complementary census of the high-redshift metal mass density, at previous inaccessible epochs. At $z\lesssim 1$, most of the metals are observed to be captured in stars \citep{Peroux20}.

Comparing our measurements of \omg\ at $z\approx 5$ and $\approx 7$ to $\Omega_{\rm dust}$ at equivalent redshifts, we can further make the first prediction for the volume-averaged dust-to-metals (DTM) ratio at these epochs. Based on the recent simulations by GADGET3-OSAKA \citep{Aoyama18} and GIZMO-SIMBA \citep{Li19}, in addition to the results quasar DLA measurements \citep{Peroux20}, we find that $\Omega_{\rm dust}/\Omega_{Z} \approx 10\%$ at $z\gtrsim 5$, about a factor of 5 lower than the Milky Way average (${\rm DTM_{Gal}} \approx 50\%$, by mass).

\section{Conclusions} \label{sec:conc}

In an attempt to establish an independent, complementary way of inferring the total ISM metal mass of galaxies, $M_{\rm Z,ISM}$, we have derived an empirical scaling between the [\cii]$-158\mu$m line luminosity and $M_{\rm Z,ISM}$. This scaling is determined from an observational benchmark sample of galaxies at $z\approx 0$, where $M_{\rm Z,ISM}$ could be estimated directly through the metallicity and \hi\ gas mass of the galaxies, in addition to recent hydrodynamical simulations of galaxies at $z\approx 0$ and $z\approx 6$. The [\cii]-to-$M_{\rm Z,ISM}$ ratio appears universal across redshifts and constant through more than two orders of magnitude in metallicity, with a ratio of $\log  (M_{\rm Z,ISM}/M_\odot) = \log (L_{\rm [CII]}/L_\odot) - 0.45$ (and 0.4 dex scatter). 

We applied this calibration to recent high-$z$ ($z\gtrsim 2$) surveys for the [\cii] line emission from main-sequence galaxies reaching well into the epoch of reionization at $z\approx 8$. We derived ISM metal masses in the range of $M_{\rm Z,ISM} = 2\times 10^{7} - 10^{9}\,M_\odot$ and found that the metal-to-stellar mass of these galaxies increases with increasing redshift. This ratio effectively describes how galaxies are increasingly efficient at retaining the produced stellar metal yield at higher redshifts, suggesting that most of the produced metals at early cosmic epochs are confined to the ISM of galaxies. This has potential important implications for outflow processes at these redshifts, which may be less substantial than previously reported.

Using the same [\cii]-to-$M_{\rm Z,ISM}$ calibration and recent estimates of the [\cii] luminosity density at $z\approx 5$ and $\approx 7$, we further placed indirect constraints on the cosmological metal mass density \omg\ at these redshifts. We found that these estimates were consistent with predictions of the total metal yield from stars, based on a recent empirical parametrization of the stellar mass density. Our measurements were therefore able to account for the total expected metal budget at these redshifts, indicating that, on average, most of the metals produced from stellar explosions are still confined to the local ISM of these galaxies. At lower redshifts, $z\approx 0-2$, most of the metals are found in other forms, predominantly stars \citep{Peroux20,Sanders23a}. Further comparing our measurements of \omg\ to $\Omega_{\rm dust}$ measured from quasar DLAs, and with recent simulations, we found that $\Omega_{\rm dust}/$\omg $\approx 10\%$ at $z\gtrsim 5$, a factor of $\approx 5$ lower than the Galactic average. 

In the near-future, the {\em James Webb Space Telescope} (JWST) will be able to routinely enable measurements of the metallicity of galaxies well into the epoch of reionization at $z\gtrsim 6,$ as previously demonstrated by the early release science data \citep[see e.g.,][]{Trump23,Schaerer22,Rhoads23,Curti23,ArellanoCordova22,Brinchmann23,Heintz23a} and the recent results from the Cosmic Evolution Early Release Science (CEERS) survey \citep{Finkelstein23}, see \citet{Heintz23b,Nakajima23,Fujimoto23,Sanders23b}. In combination with the gas masses inferred from ALMA observations through proxies such as [\cii] \citep[e.g.,][]{Heintz22} and the far-infrared continuum revealing the dust content \citep{Inami22,Dayal22}, it will soon be possible to directly measure the ISM metal mass and the DTM ratio of galaxies during the earliest cosmic epochs, as recently demonstrated in the case-study by \citet{Heintz23a}. This was enabled by combining ALMA and JWST observations of a lensed galaxy at $z\approx 8.5$. 

\section*{Acknowledgements}

We would like to thank the referee for a carefully reviewing this paper and for their suggestions that greatly improved the presentation of the results in this work.
K.E.H. acknowledges support from the Carlsberg Foundation Reintegration Fellowship Grant CF21-0103.
The Cosmic Dawn Center (DAWN) is funded by the Danish National Research Foundation under grant No. 140.

%\section*{Data availability statement} 

Source codes for the figures and tables presented in this manuscript are available from the corresponding author upon reasonable request.\\


\bibliographystyle{aa} 
\bibliography{ref}

\end{document}